\UseRawInputEncoding
\documentclass[reprint,prc,amsmath,amssymb,superscriptaddress,noffotinbib,preprintnumbers]{revtex4-1}
\usepackage{url,color,colordvi,epsfig,pstricks}
\usepackage{hyperref}
\usepackage{cleveref}
\usepackage{graphicx,bm,dcolumn}
\usepackage{natbib}
\usepackage{amsmath}
\usepackage{mathrsfs}
\hypersetup{
    colorlinks=true, 
    linktoc=all, 
    linkcolor=blue, 
    citecolor=blue}
    
\begin{document}
\preprint{FERMILAB-CONF-24-0935-PPD-T}

\title{Low-Energy Neutrino-Nucleus Scattering and New Physics}

\author{S.~Carey}\email{samcarey@wayne.edu}
\affiliation{Department of Physics and Astronomy, Wayne State University, Detroit, Michigan 48201, USA}
\affiliation{Fermi National Accelerator Laboratory, Batavia, Illinois 60510, USA}

\author{V.~Pandey}\email{vpandey@fnal.gov; Contributions to the Proceedings of the 25th International Workshop on Neutrinos from Accelerators (NuFact 2024), argonne National Laboratory, IL, USA, September 2024.}
\affiliation{Fermi National Accelerator Laboratory, Batavia, Illinois 60510, USA}

\begin{abstract}
The interactions of low-energy neutrinos with nuclei provide a unique window to explore various Standard Model (SM) and Beyond the Standard Model (BSM) processes. In particular, the recent observation of coherent elastic neutrino-nucleus scattering (CEvNS), predicted over five decades ago, has generated significant interest across disciplines. With its high cross section and suitability for compact detectors, particularly with stopped pion neutrinos, CEvNS offers a powerful probe for light, weakly coupled new physics. Ongoing global experimental efforts now aim to leverage CEvNS to test SM predictions and search for BSM signals, where deviations in event rates or spectra could reveal new physics. We present here an estimate of the number of recoil events obtained from CEvNS using the current and upcoming liquid argon based experiments. Furthermore, the event rate due to the inclusion of neutrino magnetic moment is also discussed. 
\end{abstract}

\maketitle



\section{Introduction}\label{sec:intro}

Coherent elastic neutrino-nucleus scattering (CEvNS) occurs when low-energy neutrinos scatter off the atomic nucleus as a whole via the weak neutral current, with the initial and final states of the nuclear target being indistinguishable. This process leads to a coherent contribution from all nucleons, enhancing the cross-section roughly proportional to the square of the number of neutrons in the target nucleus. The single observable from CEvNS is a recoiling nucleus, producing signals in the keV to tens of keV energy range.

CEvNS was first observed by the COHERENT experiment at the Spallation Neutron Source (SNS) using a sodium-doped CsI[Na] scintillator \cite{Akimov:2017}, decades after its prediction \cite{Freedman:1973yd}. This breakthrough not only completed the SM description of neutrino interactions with nucleons and nuclei but also opened new avenues to study weak interaction parameters, nuclear form factors, and neutrino properties beyond the SM. CEvNS experiments enable the exploration of non-standard interactions (NSI), electromagnetic neutrino properties, sterile neutrinos, new light mediators, and dark matter. They also provide crucial insights for understanding neutrino cross sections, enhancing the robustness of neutrino oscillation experiments and their interpretations. Despite its large cross-section, the small momentum transfer involved in CEvNS presents experimental challenges, hence the significance of the COHERENT experiment’s achievement in astrophysics, nuclear and particle physics.

\begin{figure}[h]
	\centering
	\includegraphics[width=1\linewidth]{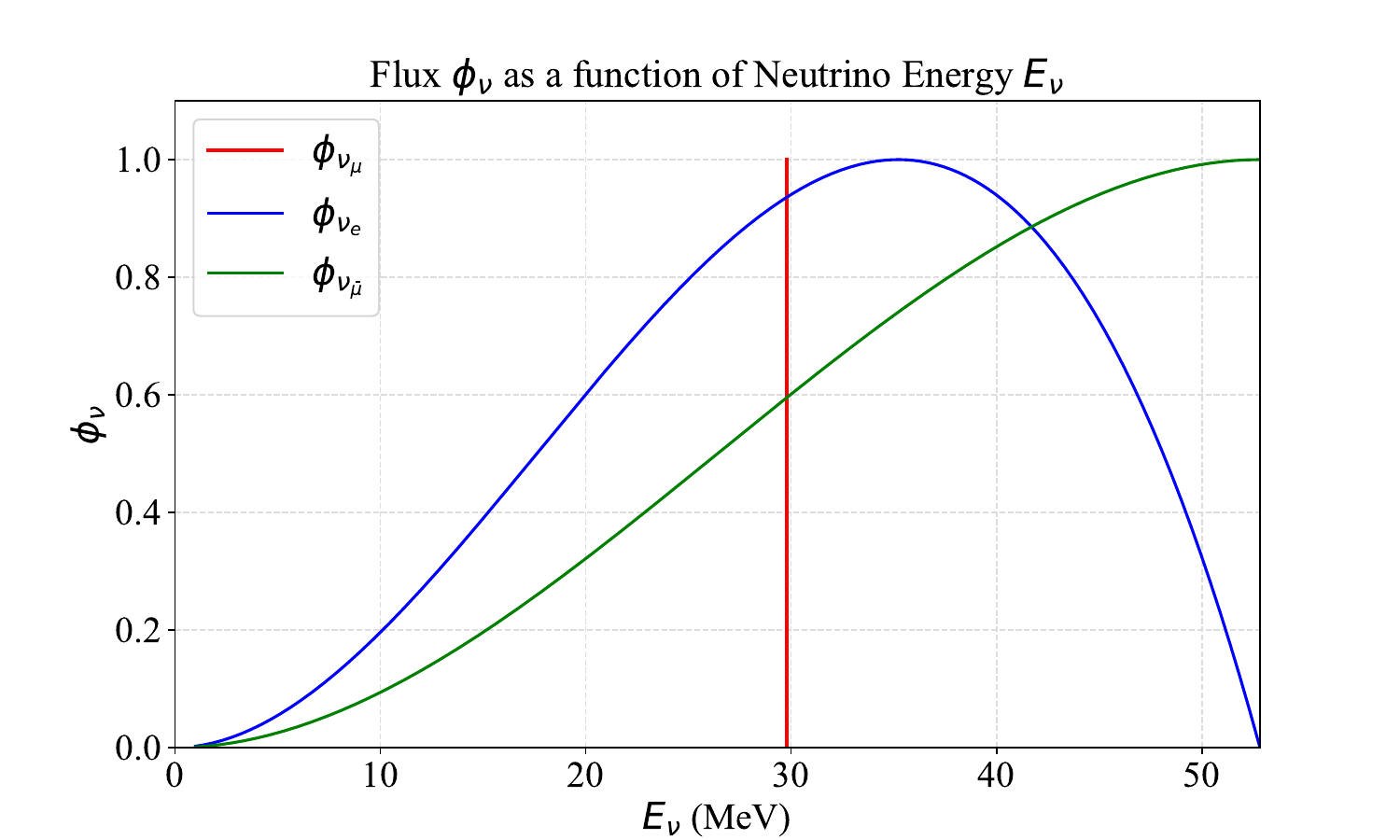}
	\caption{The standard pion decay-at-rest neutrino spectrum with 29.8 MeV monoenergetic $\nu_\mu$ while $\nu_e$ and $\bar{\nu}_\mu$ energies extending up to $m_\mu/2$.}
	\label{fig:flux}
\end{figure}

Neutrinos from stopped pion sources, with energies in the tens of MeV range, are nearly optimal for studying CEvNS. This energy range provides a balance where the CEvNS rate is sufficiently high, and the nuclear recoil energies provide detectable signals above thresholds. To date, CEvNS has only been observed using decay-at-rest (DAR) sources, making DAR neutrinos the primary focus of this discussion. The well-known energy spectrum of DAR neutrinos, shown in Fig.~\ref{fig:flux}, features a 29.8 MeV monoenergetic  $\nu_\mu$, while the $\nu_e$s and $\bar{\nu}_\mu$s energies range up to $m_\mu/2$. Additionally, this neutrino flux exhibits a characteristic timing distribution, with a \textit{prompt} $\nu_\mu$ signal followed by \textit{delayed} $\nu_e$ and $\bar{\nu}_\mu$, this pulsed time structure is a critical tool for effective background suppression.

The remainder of this article is organized as follows. Section \ref{sec:cross_section} briefly describes a formalism for calculating the CEvNS tree-level cross-section. The section also gives a brief description of inelastic neutrino scattering at a few tens of MeV. In Section \ref{sec:analysis}, the CEvNS event rates for the current and proposed liquid argon based experiments are discussed. The implications of these experiments for investigating the magnetic moment of neutrinos are also discussed in this section. Finally, we present the summary in section \ref{sec:conclusions}.


\section{Neutrino-Nucleus Scattering Cross Section}\label{sec:cross_section}

Neutrinos with tens of MeV energy can interact with detector nuclei through either (i) neutral current coherent elastic neutrino-nucleus scattering, resulting in keV-scale nuclear recoil signatures, or (ii) inelastic scattering via charged current (CC, for $\nu_e$ only) or neutral current (NC, for all flavors), producing MeV-scale energy signatures~\cite{Pandey:2023arh,VanDessel:2020epd}. The cross section for these processes is proportional to the matrix amplitude
\begin{equation}
\overline{\sum}_{fi}\left| \mathcal{M} \right|^2 = \frac{G_F^2}{2}L_{\mu\nu}W^{\mu\nu}
\end{equation}
where $G_F$ is the Fermi coupling constant, $L_{\mu\nu}$ is the lepton tensor that can be easily computed using lepton kinematics. The whole nuclear dynamics is encoded in the nuclear tensor $W^{\mu\nu}$ given as
\begin{equation}
W^{\mu\nu} = \overline{\sum}_{fi} (\mathcal{J}^{\mu}_{N})^\dagger \mathcal{J}^{\nu}_{N}
\end{equation}
where $\mathcal{J}_{N}$ represents the nuclear current amplitude. The summation symbols in these expressions represent the summation over initial polarizations and the averaging over final polarizations. 

In the CEvNS case, the nuclear current transition amplitudes are computed from the ground states ($\Phi_\textrm{0}$) to ground state ($\Phi_\textrm{0}$) transition 
\begin{equation}
\mathcal{J}^{\mu}_{N} = \langle \Phi_\textrm{0} | \widehat{J}^\mu | \Phi_\textrm{0} \rangle 
\end{equation}
and the differential cross section can then be expressed as a function of the nuclear recoil energy $T~(=Q^2/(2M_A)=E_\nu-E_\nu')$, as follows:
\begin{equation}\label{Eq:cevns_xs}
\frac{\mathrm{d}\sigma}{ \mathrm{d}T} = \frac{G^{2}_{F}}{\pi} M_{A} \left(1-\frac{T}{E_{\nu}}-\frac{M_A T}{2 E^2_\nu}\right)~\frac{Q^2_{W}}{4}~F_{W}^2(Q^2)
\end{equation}
where $E_\nu$ is the energy of the incoming neutrino, $M_A$ is the target nuclear mass, and $Q_W$ is the tree-level weak nuclear charge written as:
\begin{equation}\label{eq:weakcharge}
Q^{2}_{W} = [g_p^V Z+g_n^V N]^2 = [(1-4\sin^2\theta_\text{W}) Z-N]^2
\end{equation}
with coupling constants $g_n^V = -1$ and $g_p^V = (1-4\sin^2\theta_\text{W})$. $N$ and $Z$ are the nucleus' neutron and proton number, and $\theta_W$ is the weak mixing angle. We use $\sin^2\theta_\text{W}$ = 0.23857, the value that is valid at low momentum transfers. The entire nuclear dynamics is encoded in the elastic form factor, $F_{W}^2(Q^2)$. The maximum recoil kinetic energy of the process is limited by the kinematics of the elastic scattering
\begin{equation}
T_{\rm max} = \frac{E_\nu}{1 + M_A/(2E_\nu)}.
\end{equation}

For tens of MeV incident neutrino energies and for medium-sized nuclei, the recoil energy amounts to several tens of keV, the only experimental signature of the CEvNS process. The scattering is coherent in the sense that the nuclear structure physics that enters the cross section indeed scales with $Z$ and $N$. The form factor can be reasonably approximated by several different functional forms, such as Helm \cite{Helm:1956}, that are widely used in the CEvNS community. The simplest way of them is to denote nucleon form factors as Fourier transforms of nucleon densities:
\begin{equation}\label{Eq:Fn}
F_N(Q^2) = \frac{4\pi}{N} \int dr~r^2~\frac{\sin(Qr)}{Qr}~\rho_N(r)
\end{equation}
where $\rho_N(r)$ is the nucleon density distribution and the nucleus is considered to be spherically symmetric. The uncertainty on the CEvNS cross section is driven by the uncertainty on the neutron distribution inside the nucleus and amounts to a few percent~\cite{VanDessel:2020epd, Tomalak:2020zfh}.

\begin{figure}[h]
	\centering
	\includegraphics[width=1\linewidth]{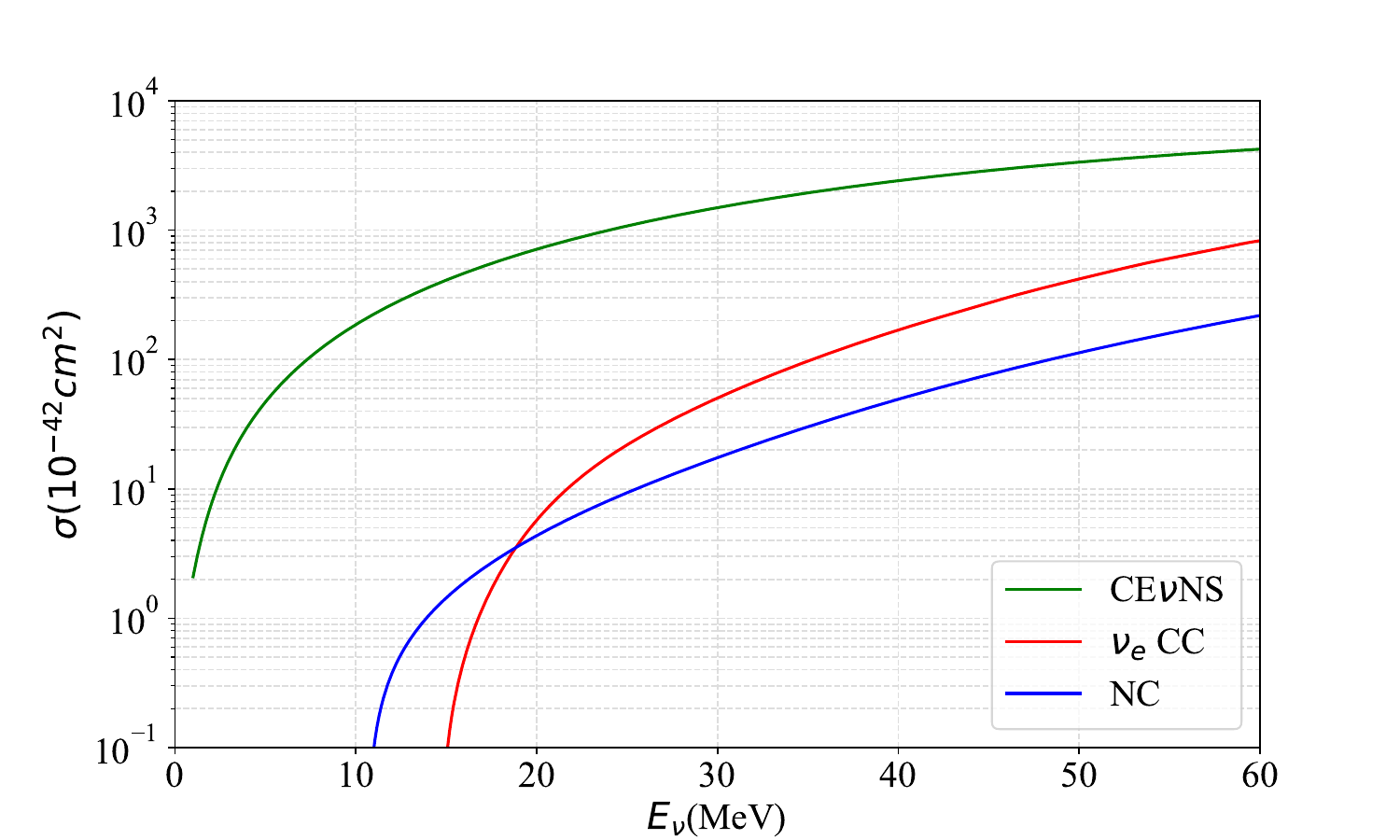}
	\caption{Comparison of the CEvNS cross section with the CC and NC inelastic scattering cross sections on argon.}
	\label{fig:xsec}
\end{figure}

In the case of inelastic scattering, the neutrino excites the target nucleus to a low-lying nuclear state, which then de-excites by emitting gamma rays or ejecting nucleons. The nuclear transition amplitude are computed between an initial $| \Phi_\textrm{0} \rangle$ and final a $| \Phi_\textrm{f} \rangle$ state:
\begin{equation}\label{eq:current}
\mathcal{J}^\mu_{N} = \langle \Phi_\textrm{f} | \hat{J}^\mu | \Phi_\textrm{0} \rangle
\end{equation}
and the differential cross section of this process can be written as 
\begin{equation}\label{eq:xsec}
\begin{aligned}
\frac{\mathrm{d}^3\sigma}{\mathrm{d}\omega\mathrm{d}\Omega} =& \frac{G^2_F}{4\pi^2} ~\cos\theta_c~E_f k_f \\
&\times \left( v^{\mathcal{M}} R^{\mathcal{M}} + v^{\mathcal{L}} R^{\mathcal{L}} + v^{\mathcal{ML}} R^{\mathcal{ML}} \right.  \\
& + \left. v^{T} R^{T} + h v^{TT} R^{TT} \right)
\end{aligned}
\end{equation}
where $\cos\theta_c$ is the Cabibbo angle included for the CC reaction (this factor is absent in the NC case), $\omega$ is the energy transferred to the nucleus. The energy and momentum of the outgoing lepton is denoted by $E_f$ and $k_f$, respectively. The $v$--factors are leptonic functions and $R$--factors are the nuclear response functions. The indices $L$ and $T$ correspond to longitudinal and transverse contributions. At tens of MeV energies, the CEvNS cross section is orders of magnitude larger than the inelastic cross section, as shown in Fig.~\ref{fig:xsec}.

At the stopped-pion neutrino sources, the CC reactions are accessible only for $\nu_e$s while NC reactions are available for all neutrino types. The inelastic scattering process is subject to details of underlying nuclear structure physics and is less well understood. These are however important to study at the stopped-pion sources given their energies overlap with the energies of neutrinos from core-collapse supernova \cite{VanDessel:2020epd, Pandey:2023arh, DUNE:2023rtr}.


\section{Analysis and Results}\label{sec:analysis}

Several liquid argon based experimental programs have been established or are underway to probe CEvNS and new physics signals using stopped-pion neutrino sources \cite{Akimov:2021, Aguilar-Arevalo:2022, Aguilar-Arevalo:2023}. In Tab.~\ref{tab:CEvNS}, we list current or proposed liquid argon based CEvNS detectors at the stopped pion sources. Since the uncertainties on the SM-predicted CEvNS cross section is relatively small, measurements of CEvNS allow tests of SM weak physics or probing new physics signals. \\

\begin{table}[htbp]
\centering
\begin{tabular}{|c|c|c|c|c|c|}
\hline
Experiment & Mass & Distance from source & Dates \\
 & (kg) & (m) & \\
\hline \hline 
CENNS10  & 24  & 27.5  & 2017 - \\ 
    (ORNL)  &  &   &  \\
\hline 
CENNS750  & 610  & 27.5  & proposed \\ 
    (ORNL)  &  &   &  \\
    \hline
CCM & 10,000  & 20.0   & 2019 -   \\ 
(LANL) &  &   &  \\ 
\hline
PIP2-BD at F2D2 & 100,000  &  20.0  & proposed   \\ 
(FNAL) &  &   &  \\ 
\hline
\end{tabular}
\caption{Current or proposed CEvNS experiments with liquid argon detectors at the stopped pion sources.}
\label{tab:CEvNS}
\end{table}

The differential event rate for different neutrino flavors as a function of recoil energy is
\begin{eqnarray}\label{Eq:cevns_rate}
    \frac{dN}{dT} = \mathcal{N}\sum_{i = \nu_\mu, \nu_e, \bar\nu_\mu} \int_{E_\nu^{min}}^{m_\mu/2} dE_\nu \frac{d\phi_{\nu}}{dE_\nu} \frac{d\sigma}{dT}
\end{eqnarray}
where, the factor $\mathcal{N}$ is defined in terms of the detector mass, distance (from the source), protons-on-target, and the neutrino yield per proton. The interaction with argon nucleus is parametrized by the Helm form factor \cite{Helm:1956} in Eq.~\ref{Eq:cevns_xs}. In the Helm approach~\cite{Helm:1956}, the form factor is expressed as: 
\begin{equation}
F_{\text{Helm}}(Q^2) = \frac{3 j_1(QR_0)}{QR_0} e^{-Q^2s^2/2}
\end{equation}
where $j_1(x) = \sin(x)/x^2 - \cos(x)/x$ is a spherical Bessel function of the first kind. $R_0$ is an effective nuclear radius given as: $R_0^2 = (1.23 A^{1/3} - 0.6)^2 + \frac{7}{3} \pi^2 r_0^2 - 5 s^2$ with $r_0$ = 0.52 fm and the folding width $s$ = 0.9 fm, fitted to muon spectroscopy and electron scattering data \cite{Fricke:1995zz}. 

\begin{figure}[h]
	\centering
	\includegraphics[width=1\linewidth]{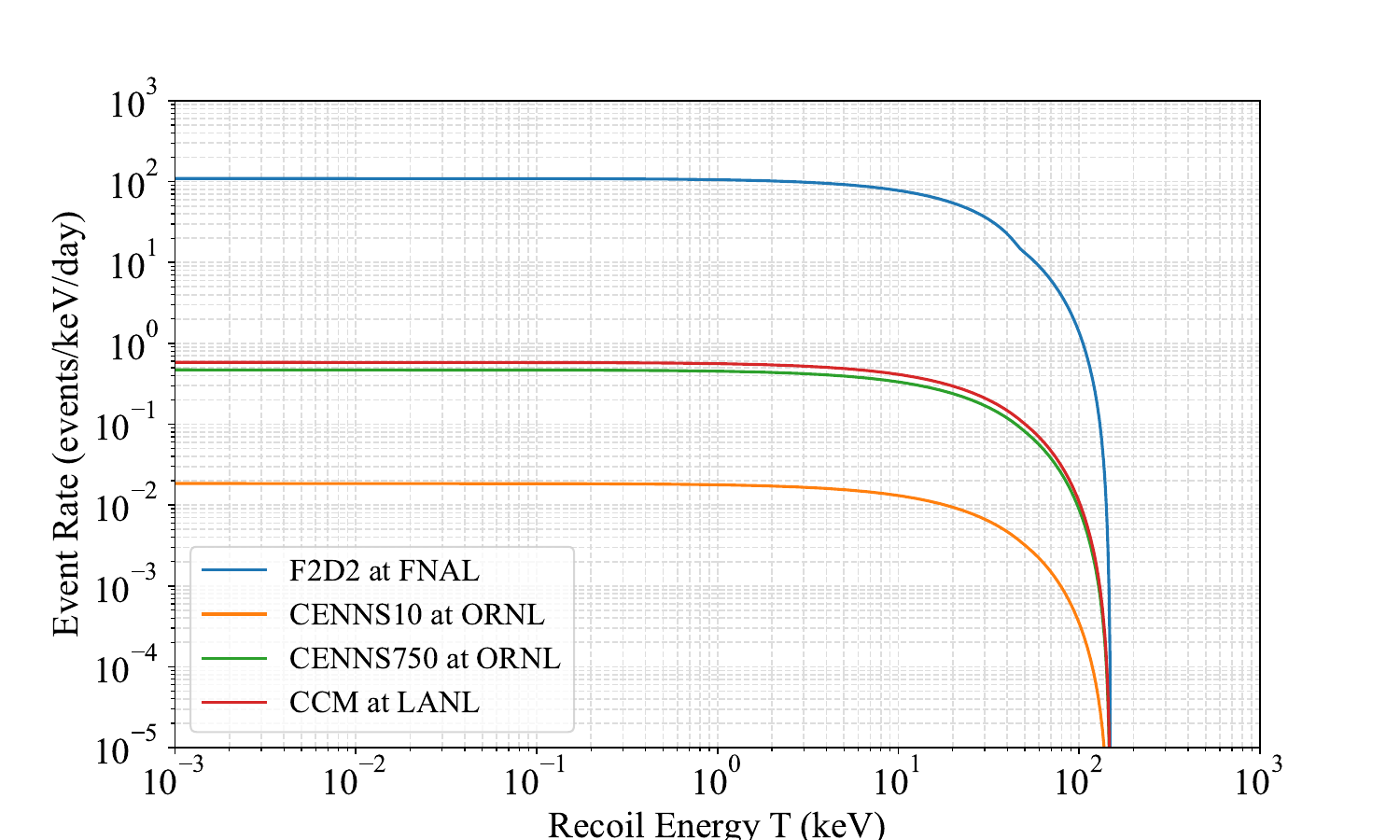}
	\caption{Expected CEvNS event rates as a function of nuclear recoil energy for different liquid argon detectors.}
	\label{fig:Rate_Fluxes}
\end{figure}

The number of nuclear recoil events obtained per day for the different experiments is shown in Fig.~\ref{fig:Rate_Fluxes}. The kink at roughly 50 keV is due to the endpoint of the recoils caused by the prompt $\nu_\mu$ flux. Any deviation from the rate obtained from the standard model prediction could indicate new contributions to the interaction cross-section. The new contributions could just be coming from the defined parameters like weak mixing angle, neutron radius or from some new source.

The existence of massive neutrinos from the discovery of neutrino oscillations is a clear indication of the physics beyond the standard model. An effective way to look for this is to probe the electromagnetic property of the neutrino. One such property, the magnetic moment arises naturally from the neutrino electromagnetic vertex \cite{Giunti:2014}. 

\begin{figure}[h]
	\centering
	\includegraphics[width=1\linewidth]{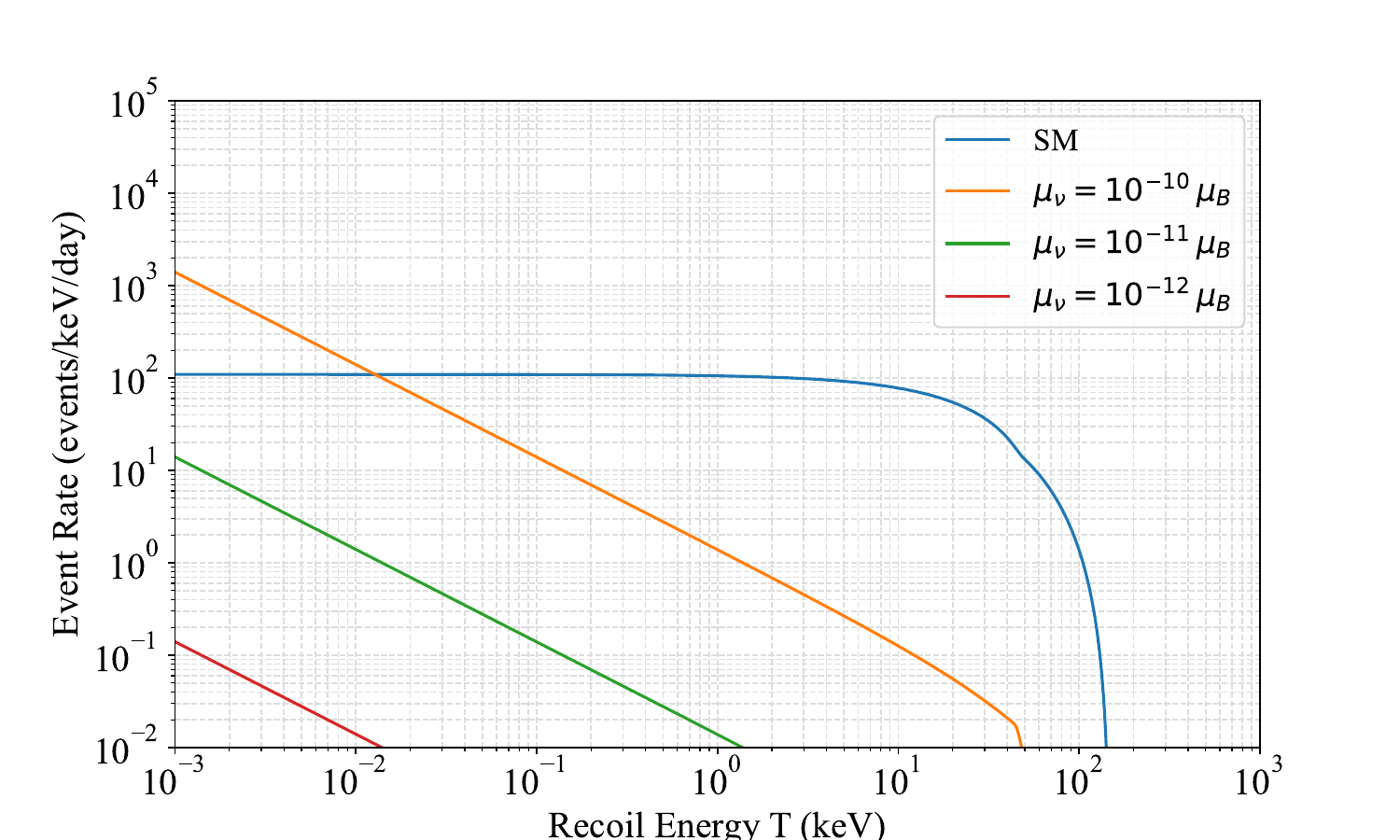}
	\caption{Expected nuclear recoil rate from CEvNS due to neutrino magnetic moment. The rates shown are due to flux from the PIP2-BD at F2D2 for the SM contribution (solid blue) as well as the contributions due to different values of the neutrino magnetic moment $\mu_\nu$.}
	\label{fig:Rate_MM}    
\end{figure}

In the extended electroweak theory that accounts for neutrino masses, the neutrino magnetic moment is given by \cite{pdg:2024}:
\begin{eqnarray}
    \mu_\nu = 3.2 \times 10^{-19} \left(\frac{m_\nu}{\text{eV}}\right) \mu_B.
\end{eqnarray}
Even though this value of the magnetic moment is very small and almost unobservable, but for more general models, there is no longer a proportionality between neutrino mass and its magnetic moment. Direct dark matter detection experiments \cite{Aprile:2022} and COHERENT \cite{Coloma:2022} has put bounds on the magnetic moment of the neutrino of the order $10^{-10} \mu_B$. The nuclear recoil event rates due to the addition of neutrino magnetic moment are shown in Fig.~\ref{fig:Rate_MM}. The differential cross section in the presence of a neutrino magnetic moment is given by the expression
\begin{eqnarray}\label{eq:MM}
    \left( \frac{d\sigma}{dT} \right)_{\text{EM}} = \frac{\pi \alpha^2 \mu_\nu^2 Z^2}{m_e^2} \left( \frac{1}{T} - \frac{1}{E_\nu} + \frac{T}{4 E_\nu^2} \right) F_{\text{ch}}^2 (Q^2)
\end{eqnarray}

Where $\alpha$ is the EM coupling constant, $F_{\rm{ch}}$ is the charge form factor of the argon nucleus and $m_e$ is the electron mass. This factor only adds incoherently to the SM cross-section, while the interference terms vanish due to the required spin-flip. As observed in Fig.~\ref{fig:Rate_MM}, for a sufficiently low detection threshold, the inclusion of magnetic moment results in an enhancement of the recoil spectrum. 

\section{Summary}\label{sec:conclusions}

The low-energy neutrino-nucleus scattering, especially the coherent elastic scattering process, provides a unique window to explore various SM and BSM physics scenarios. Leveraging orders of magnitude higher CEvNS cross section, new physics can be searched with relatively small detectors. We estimate an improvement in the number of nuclear recoil events for upgrades to current and future CEvNS experiments compared to the previous experiments. These experiments, including those at ORNL, LANL, and Fermilab, aim to test SM predictions and search for deviations that could indicate new physics, such as the neutrino EM property, non-standard interactions or new light mediators.

We also examine the effect of neutrino magnetic moments on nuclear recoil rates as a potential signature of new physics. The inclusion of a neutrino magnetic moment ($\mu_\nu$) enhances the recoil spectrum, particularly at low energies. We show how the event rates are modified for different values of $\mu_\nu$ compared to the SM prediction. This enhancement provides a potential method for probing the electromagnetic properties of neutrinos and constraining the neutrino magnetic moment.

\acknowledgments 
We thank Jacob Zettlemoyer for providing neutrino flux of F2D2 at Fermilab. V.P. is grateful to the organizers of the NuFACT 2024 workshop for the invitation and hospitality. S.C. would like to acknowledge the support of U.S. Department of Energy grant DE-SC0007983 and the Visiting Scholars Award Program of the Universities Research Association. This manuscript has been authored by Fermi Research Alliance, LLC under Contract No. DE-AC02-07CH11359 with the U.S. Department of Energy, Office of Science, Office of High Energy Physics.


\bibliography{references}


\end{document}